\documentclass[journal,transmag]{IEEEtran}
\usepackage{cite}

\usepackage[final]{graphicx}
\ifCLASSINFOpdf
\else
\fi

\ifCLASSOPTIONcaptionsoff
 \renewcommand{\caption}[2][\relax]{\MYoriglatexcaption[#2]{#2}}
\fi
\begin{document}
%
\title{Cybertwin-enabled 6G Space-air-ground Integrated Networks: Architecture, Open Issue, and Challenges}


\author{
	\IEEEauthorblockN{
Zhisheng Yin\IEEEauthorrefmark{1},~\IEEEmembership{Member,~IEEE},
Tom H. Luan\IEEEauthorrefmark{1},~\IEEEmembership{Senior Member,~IEEE},
Nan Cheng\IEEEauthorrefmark{2},~\IEEEmembership{Member,~IEEE}, 
\\ Yilong Hui\IEEEauthorrefmark{2},~\IEEEmembership{Member,~IEEE},
and Wei Wang\IEEEauthorrefmark{3},~\IEEEmembership{Member,~IEEE} 
}
\IEEEauthorblockA{\IEEEauthorrefmark{1}State Key Lab. of ISN and School of Cyber Engineering, Xidian University, Xi'an, 710071, China}
\IEEEauthorblockA{\IEEEauthorrefmark{2}State Key Lab. of ISN and School of Telecommunications Engineering, Xidian University, Xi'an, 710071, China}
\IEEEauthorblockA{\IEEEauthorrefmark{3}College of Electronic Information Engineering,
	Nanjing University of Aeronautics and Astronautics, Nanjing, 211106, China}
%
%
\thanks{This work was supported by National Natural Science Foundation of China (No. 62071356). Corresponding author: Tom H. Luan and N. Cheng (email: tom.luan@xidian.edu.cn; dr.nan.cheng@ieee.org).}
%
}

\markboth{Journal of \LaTeX\ Class Files,~Vol.~14, No.~8, August~2022}%
{Shell \MakeLowercase{\textit{et al.}}: Bare Demo of IEEEtran.cls for IEEE Transactions on Magnetics Journals}
%



\IEEEtitleabstractindextext{%
\begin{abstract}
  Space-air-ground integrated network (SAGIN) is considered as a core requirement in emerging 6G networks, which integrates the terrestrial and non-terrestrial networks to reach the full network coverage and ubiquitous services. To envision the ubiquitous intelligence and the deep integration in 6G SAGIN, a paradigm of cybertwin-enabled 6G SAGIN is presented in this paper. Specifically, a cybertwin-enabled SAGIN architecture is first presented, where a novel five-dimension digital twin (DT) model is presented. Particularly, three categories of critical technologies are presented based on the cybertwin of SAGIN, i.e., cybertwin-based multi-source heterogeneous network integration, cybertwin-based integrated cloud-edge-end, and cybertwin-based integrated sensing-communication-computing. Besides, two open issues in the cybertwin-enabled SAGIN are studied, i.e., the networking decision and optimization and the cybertwin-enabled cross-layer privacy and security, where the challenges are discussed and the potential solutions are directed. In addition, a case study with federal learning is developed and open research issues are discussed. 
\end{abstract}

\begin{IEEEkeywords}
Cybertwin, 6G, SAGIN, digital twin.
\end{IEEEkeywords}}

\maketitle

\IEEEdisplaynontitleabstractindextext

%
\IEEEpeerreviewmaketitle

\section{Introduction}
%
%
%
%
\IEEEPARstart{S}{pace}-air-ground integrated network (SAGIN) enables full network coverage range, which can increase the network capability, provide seamless connectivity, and enhance network performance \cite{Zhou2019}. In the emerging 6G networks, the particular features of intelligence, complexity, dynamics, and customization are emphatically concerned, aiming to cover full service scenarios and full coverage. On the exploration of developing 6G networks, SAGIN shows a considerable potential and has attracted extreme attentions in academia and industry \cite{you2021towards}. However, the key technologies and methods in SAGIN network service fulfillment are still under investigating.

SAGIN is known as a multi-tier network consists of space networks, aerial networks, and terrestrial networks, which are recognizably heterogeneous and hold respective advantages and challenges\cite{Liu2018}. Particularly, the space networks of satellites broadcast widely and have strong damage resistance of disasters, while the extra long-link transmission indicates long delay and the limited resource (e.g., power, computing, and transponders, etc.) at satellite manifests  low processing capability. 
For the aerial networks, high-altitude platforms (HAPs) and unmanned aerial vehicles (UAVs) can be dynamically deployed to independently provide services or assist a special area, which is flexible to self-organize and expand \cite{Ma2021}. Whereas the terrestrial networks develop faster and have relatively mature technologies, it holds appreciable capability and supports rapid-rate transmissions and lower latency with high throughput. Particularly, such multi-tier networks are dynamically heterogeneous in multi-dimensional resources and different transmission protocols. Considering the service fulfillment is significantly affected by different network protocols, dynamic link conditions, and heterogeneous resources, it renders the service fulfillment in SAGINs overwhelmingly challenging. Specifically, there exposes challenges for SAGIN to intelligently and effectively operate as a whole network, such as
\begin{itemize}
	\item \textbf{Mobility}: Low earth orbit (LEO) satellites operate under the high mobility condition since its non-geostationary characteristics, which requires the efficient communication and handover of satellite constellation to response immediately. Besides, the aerial network provider such as the UAV is also deployed with the high mobility and its opportunistic access strategy is hard to guarantee the continuous communication. Thus the challenge of mobility management also refers to the coordination between the three segments in SAGIN.  
	\item \textbf{Heterogeneity}:
	SAGIN is a typical heterogeneous network, since it is consisted of multi-tier networks with different network and communication protocols. The terminals of SAGIN are expected to be designed as multi-mode and handover among different access networks smoothly. 
	However, the different node resources have the differentiated functions, the key to reach the deep integration in SAGIN is to adequately utilize the distinguished characteristics and efficiency of such nodes for achieving the fusion with maximum efficiency.  
	\item \textbf{Latency}: Compared with the terrestrial networks, the long transmission link between satellite and the ground indicates the large latency. It is challenging to complete a task/service with low latency through satellite. Therefore, the latency-sensitive services are generally expected to be executed in one-time transmission, which needs the high reliability of communications.  
\end{itemize}
\begin{figure}[!t]
	\centering
	\includegraphics[width=0.43\textwidth]{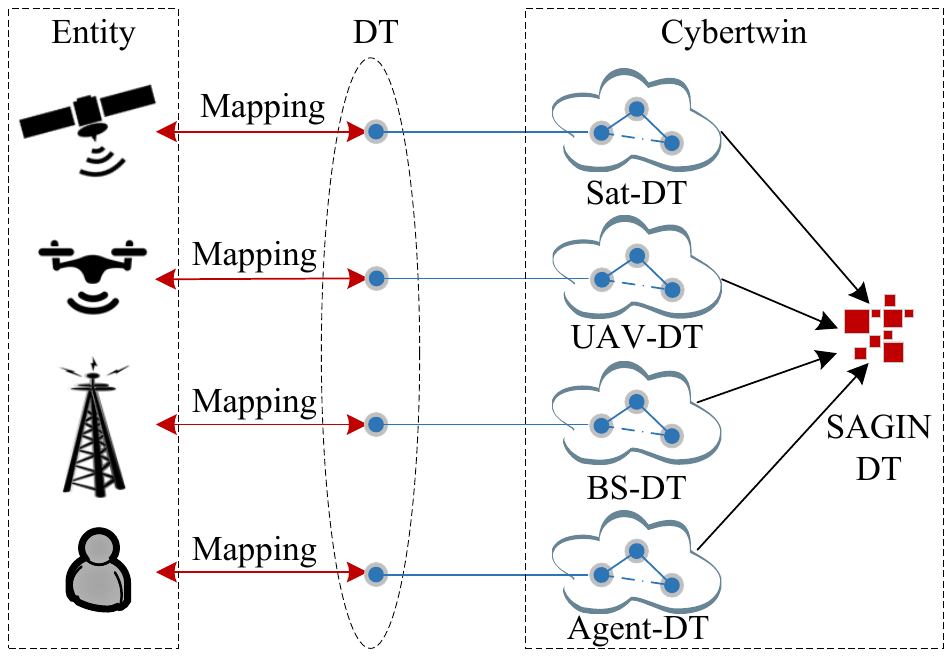}
	\caption{DT mapping model in SAGIN.}
	\label{fig1a}
\end{figure}
With the vision that 6G networks connecting the physical world to the digital world,   
the requirements of digitization and intelligence for SAGIN is an irresistable trend.
The digital twin (DT) technology provides a promising paradigm. A DT is the precise virtual copy of a physical machine or system mirrors almost every facet of a product, process or service \cite{Tao2019a,Wu2021}.
Initially, the DT concept model is first presented in \cite{Grieves2014}, which contains three main parts, i.e., physical products in real space, virtual products in virtual space, and the connections of data and information that ties the virtual and real space together.
 Driven by but not limit to the DT, a cybertwin network architecture for 6G is first presented in \cite{Yu2020,yu2019cybertwin}, where the cybertwin serves as communications assistant, network behavior logger, and digital asset owner.
Appreciated to this work, the cloud-centric network architecture and radio
access network (RAN) architecture are proactively given, and particularly resources (computing, caching, and communications resources) allocation coordinately and decoupling the control and user planes are addressed.
 Particularly, a basic DT communication model are first presented in \cite{Luan2021}, including inter-twin and intra-twin communication models, and a case study of DT system for autonomous vehicles is conducted. Inspired by the investigations of DT and cybertwin, 
 it motivates us to consider a new framework of cybertwin-driven SAGIN to decouple its complex network management and fulfill diversified services.

In this paper, a novel cybertwin-enabled SAGIN architecture to enable the digital reflection and the intelligent management of SAGIN is presented. The main work includes the following.
\begin{itemize}
\item To address the multi-tier heterogeneity and dynamic networking, we present the cybertwin-enabled multi-source heterogeneous network integration in SAGIN, where a cybertwin-enabled SAGIN resource orchestration flow is conducted to support the service-oriented resource management. 
\item To improve the networking efficiency and enhance an inherent coordination, we further depict the cybertwin-enabled integrated cloud-edge-end in SAGIN, where the integrated cloud-edge-end resource pool is established and the  cybertwin-enabled collaborations among local ends, edges, and the core cloud can be reached.
Considering the scene information and user intentions, the cybertwin-enabled integrated sensing-communication-computing model is presented to improve the differentiated service experiences.  
\item Moreover, the future research issues and the potential solutions are directed in the cybertwin-enabled SAGIN, i.e., AI-based networking decision and optimization, 
and federated learning (FL) based cross-layer privacy and security. 
In addition, a case study of model training in the cybertwin-enabled SAGIN has been conducted through a FL approach.   
\end{itemize}

\section{Cybertwin-enabled SAGIN Architecture}

To represent the physical SAGIN entity digitally, we depict a five-dimension SAGIN DT model 
to conduct the SAGIN cybertwin, which can be deployed at the core and edge clouds in the physical SAGIN. 
Particularly, we define the five-dimension SAGIN DT model and the DT mapping model is shown in Fig. \ref{fig1a}.
 \begin{figure*}[h]
	\centering
	\includegraphics[width=0.95\textwidth]{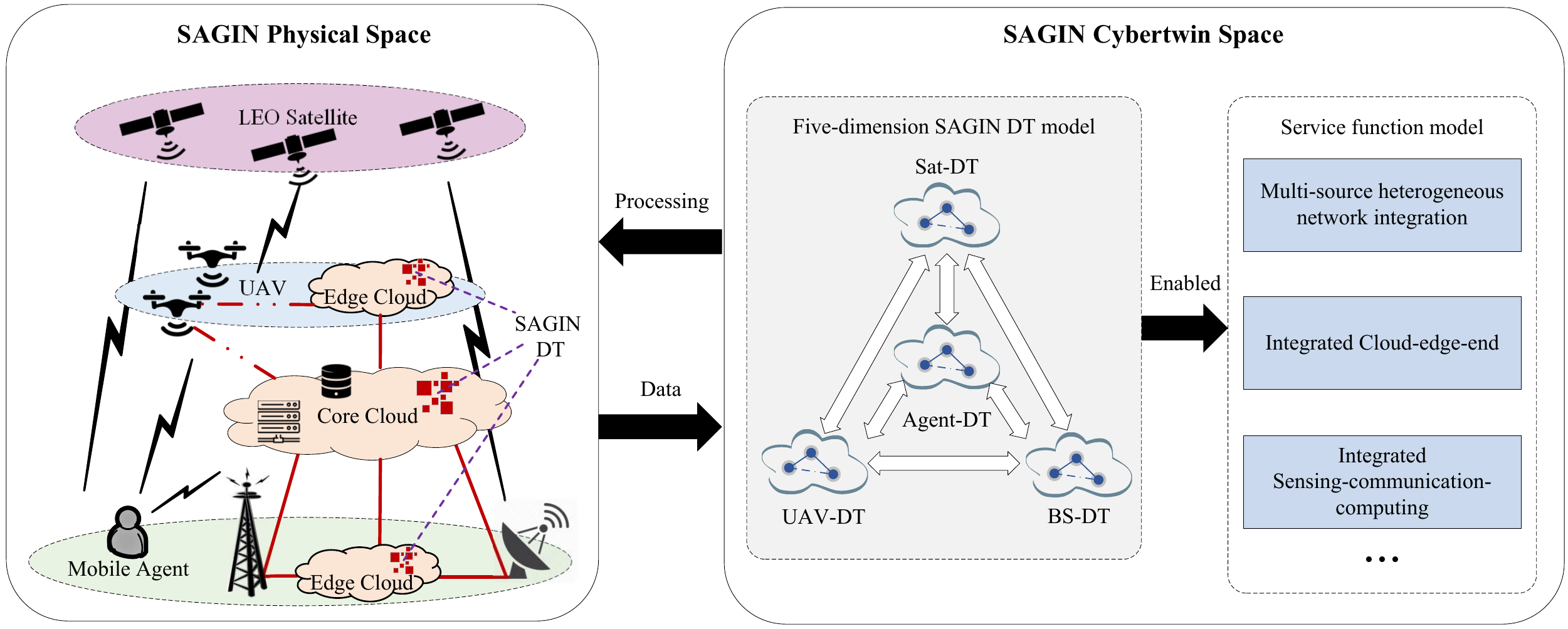}
	\caption{Cybertwin-enabled SAGIN architecture.}
	\label{SM}
\end{figure*}
\begin{itemize}
	\item \textbf{Sat-DT} represents the DT of satellites. It holds the activity information of LEO satellites, including the orbiting logger and the dynamic resource status.
	\item \textbf{UAV-DT} represents the DT of aerial UVAs. Similarly, it directs the mobility of UAVs trajectory and holds the dynamic time-varying network status.
	\item \textbf{BS-DT} represents the DT of terrestrial BSs. It manages the communication and computation resources of BS.
	\item \textbf{Agent-DT} represents the DT of mobile agent. It supervises and records the activity information of mobile agent.
	\item \textbf{Connection} between different kinds of DTs is established to exchange information crossing multi-tier RANs.
\end{itemize}

\begin{figure*}[h]
	\centering
	\includegraphics[width=0.7\textwidth]{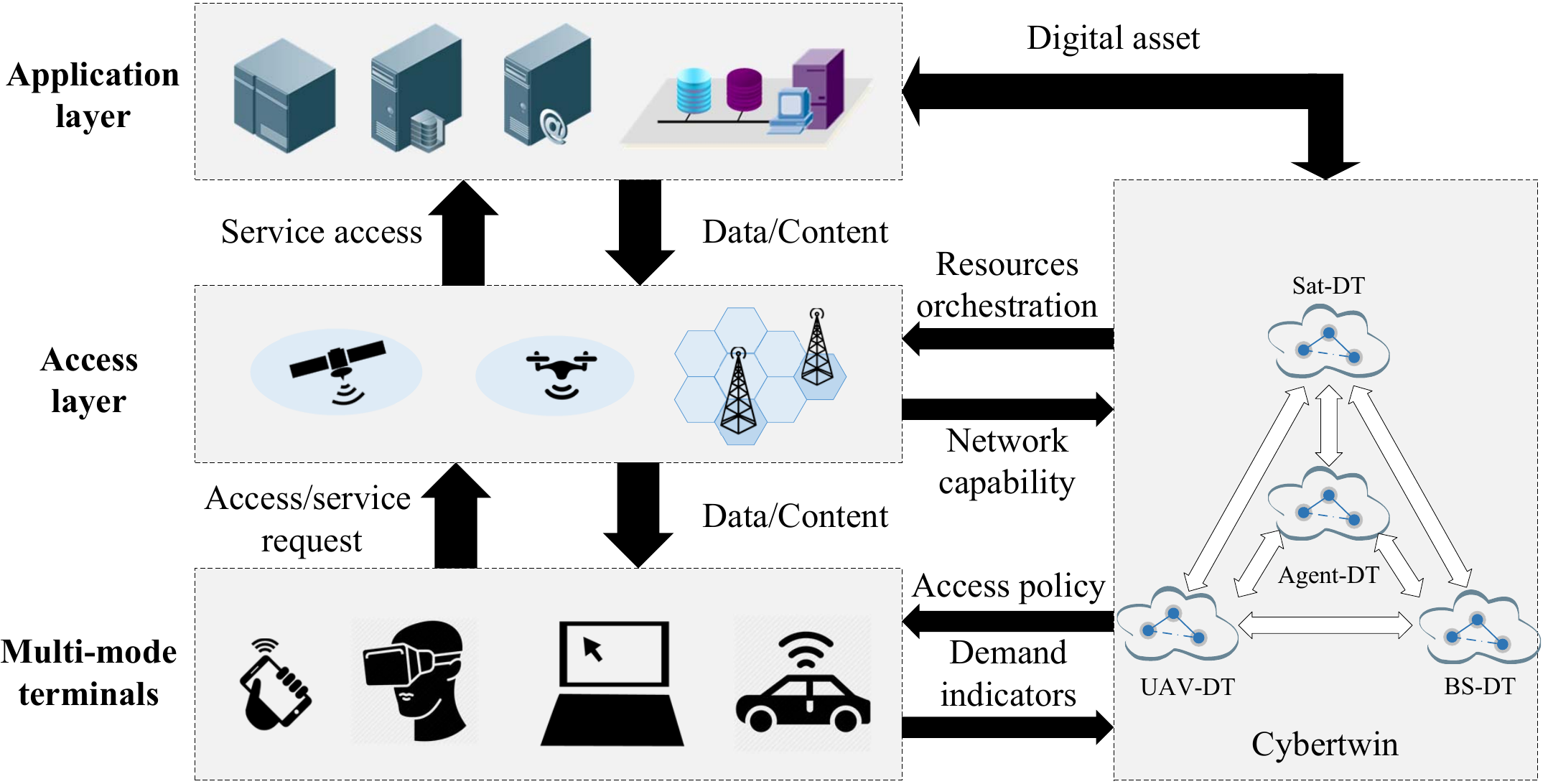}
	\caption{Cybertwin-enabled SAGIN resource orchestration flow.}
	\label{flow}
\end{figure*}
Fig. \ref{SM} shows our developed cybertwin-enabled SAGIN architecture, where
the physical space and the cybertwin space of SAGIN are included.
In the SAGIN physical space, there are three-layer connected radio access networks (RANs), i.e., LEO satellite RAN, aerial RAN, and terrestrial RAN. The mobile agents can access such three-layer RANs according to the different requirements.

With the data information collected from the SAGIN physical space, 
the physical SAGIN can be reflected veritably by the DTs in SAGIN cybertwin space and
the cybertwin-enabled service function models can be realized by the data analyzing and processing. Moreover, such service function models care be mapped into the SAGIN physical entity to configure and adjust the network operation and it the on-demand service can be directed. Particularly, we present three categories of critical technologies based on the cybertwin of SAGIN as follows, i.e., cybertwin-based multi-source heterogeneous network integration, cybertwin-based integrated cloud-edge-end, and cybertwin-based integrated sensing-communication-computing   

\begin{center}
	\textbf{Cybertwin-based Multi-source Heterogeneous Networking Integration in SAGIN}
\end{center}

SAGIN is a typical multi-source heterogeneous network and the compatibility  of such multi-mode communications and networking is crucial for the integration of multi-source heterogeneous networks in SAGIN. 
Particularly, how to integrate the heterogeneous resource efficiently and effectively is still an open issue.
Considering the increasing diverse services which hold different key performance indicators (KPIs), the service-oriented networking has attracted great attentions and it shows potential in providing flexible network service and enhanced network performance by accurately matching the network resource and the service requirements. 

Mobile agents such as multi-mode devices/terminals generally have different KPIs and they are permitted to access satellite, aerial, and terrestrial RAN selectively. 
Different service KPIs have different demands on resource, e.g., communication and computation resources. 
To improve the spectral efficiency and satisfy the quality of service (QoS) of users, we arguably consider that all the available resources in SAGIN are open to any agent and allocated by orienting on-demand. 

To achieve the deep integration of multi-source heterogeneous network in SAGIN, we present a cybertwin-enabled SAGIN resource orchestration flow as shown in  
Fig. \ref{flow}.
Particularly, the access layer is  only responsible for receiving and forwarding the access and service requests in the uplink, and delivering data/contents to users in the downlink. According to the demand indicators, cybertwin as the brain of SAGIN provides access policy and resource orchestration, which holds the global information and makes decisions on behalf of agents and physical network entity. 

\begin{figure*}[h]
	\centering
	\includegraphics[width=0.65\textwidth]{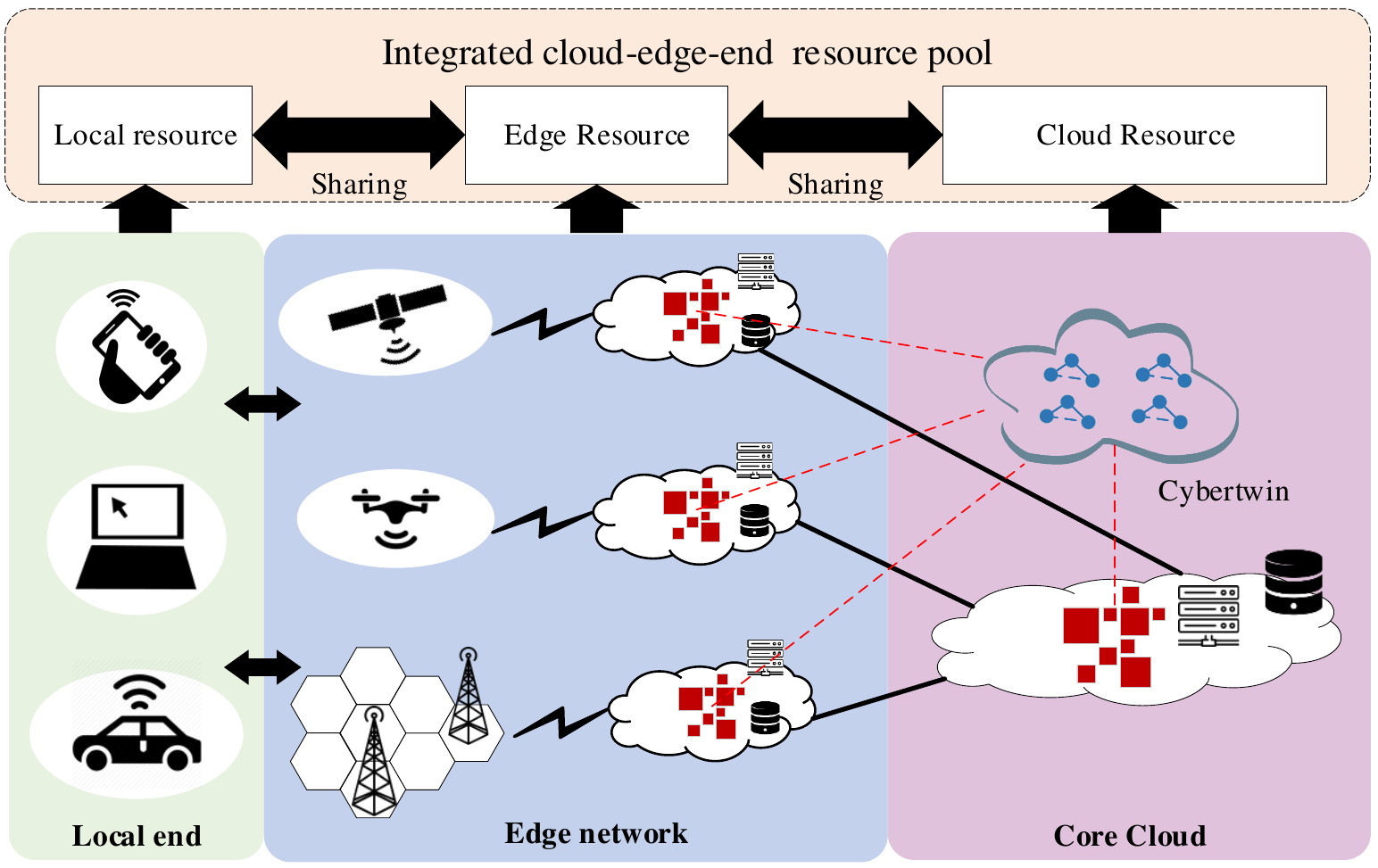}
	\caption{Cybertwin-based integrated cloud-edge-end in SAGIN}
	\label{CEE}
\end{figure*}
Moreover, in the cybertwin space, the four kinds of DTs with different numerologies are grouped in a meeting and they can exchange information with each other. Thus, the decisions of access selection and the resource allocation for users are multilaterally bargained at that DT meeting. Without loss of generality, each couple of entity and its DT are symbiotic and they can identify each other. When the service requested by an agent is being ended, the agent-DT can beforehand broadcast resource release information, which can support a precognition and preparing for new services. In addition, considering the DTs hold the behavior data logging of mobile agents, it can be converted into a digital asset for the science/project experiment and sale after removing sensitive information. 

\begin{center}
	\textbf{Cybertwin-based Integrated Cloud-edge-end in SAGIN}
\end{center}

To support diversified and differentiated services from a prospective of on-demand fulfillment, it is considerably imperative to regulate a consensus that such multi-source heterogeneous resources can be shared and collaboratively orchestrated. Benefiting from wide investigations of resource virtualization and customized network functions such as network slicing, software-defined networking (SDN), and network function virtualization (NFV) techniques\cite{Zhuang2020}, hence since it is indispensable and reasonable to conduct the integrated cloud-edge-end in SAGIN. Fig. \ref{CEE} shows our developed cybertwin-based integrated cloud-edge-end in SAGIN, it mainly consists of three-segment service and resource providers, i.e., local end, edge network, and core cloud. Particularly, although the core cloud has abundant computation resource and supports the large-scale and complex services, mobile edge computing (MEC) is proposed for decentralizing cloud services to the network edges to achieve lower latency, and such three edge networks in SAGIN can collaboratively construct the service function chain (SFC). 
Besides, the local terminals also have certain processing capability and they can also share resource with other services.
From Fig. \ref{CEE}, it can be also seen that the cybertwin unit exists in the cloud and edges and each DT of an end or service serves as the agent with full authority to maximize the best service experience for its host. 

Without loss of generality, considering a comprehensive and intricate service in SAGIN, a task offloading is generally executed when the local servers cannot process timely to reduce the energy consumption of local end and the latency. Different from traditional greedy schemes, a globally optimal offloading strategy is prepared by such DTs in cybertwin space, where the integrated cloud-edge-end resource pool is scheduled. Thus, a task workflow can be carried out through a cooperation of local end, edge network, and core cloud. Moreover, an assistance among local ends or task migration among edges with different numerologies can be also executed according the result of DT decision. Particularly, considering those services in remote areas where satellite constellation provides the RAN, task migrations to cloud or other edges can improve the service efficiency, due to the limited resource at satellites. 

\begin{center}
	\textbf{Cybertwin-based Integrated Sensing-communication-computing in SAGIN}
\end{center}

\begin{figure*}[!h]
	\centering
	\includegraphics[width=0.65\textwidth]{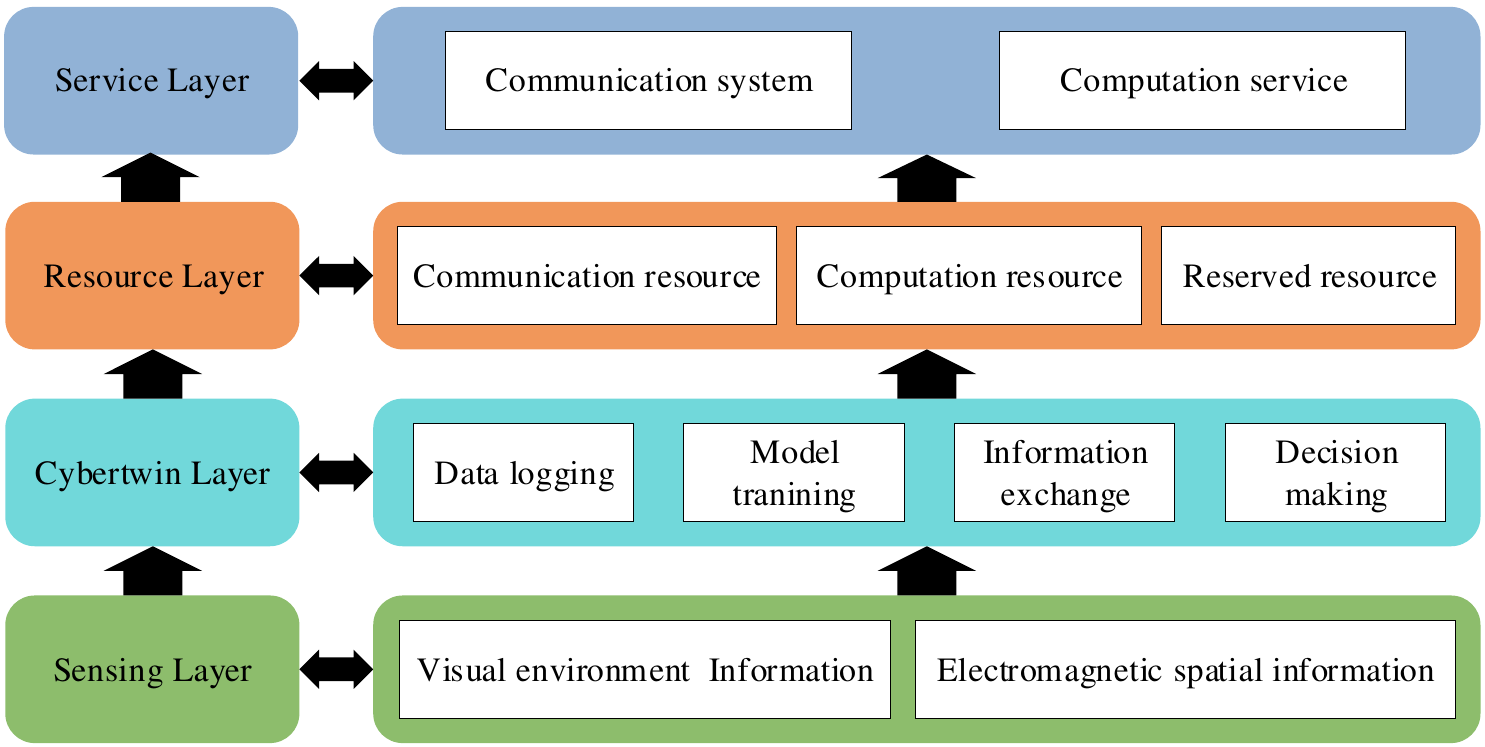}
	\caption{Cybertwin-enabled integrated sensing-communication-computing model in SAGIN}
	\label{SCC}
\end{figure*}
A cybertwin-enabled integrated sensing-communication-computing model in SAGIN is presented in Fig. 4, where the sensing layer, cybertwin layer, resource layer, and service layer are included. We aim to conceive a service-orient system model and conduct a cybertwin-centric guideline. Then the four-layer cybertwin-enabled integrated sensing-communication-computing model is introduced respectively through these four layers as follows.   
 
\textbf{Sensing layer:} Mobile terminals with intelligent sensing devices can collect visual environment information and electromagnetic spatial information through sensors. The visual environment information contains the surrounding environment and mobility of mobile agents, referring to environmental monitoring, target detection and identification, and the state parameter measurement. For instance, the road condition, traffic command information and the state of vehicles in motion can be collected by the visual sensors mounted on vehicles. The electromagnetic spatial information can reflect a state of networking and communication, where the knowledge of spectrum occupancy and channel state information etc. could be a local feedback to improve communication performance. Particularly, we assume all the mobile agents, satellite, UAV, and BS have the ability of spectrum sensing. Such sensed information is updated in real-time and gathered in cybertwin. 

\textbf{Cybertwin layer:} In the cybertwin space, the DTs hold its corresponding sensed information respectively, which can be used to analyze and predict some inertial behaviors and direct a quick response. To support  customized services, we conduct four main functions such as data logging, model training, information exchange, and decision making in the cybertwin space of SAGIN. For the function of data logging, the DT logs fresh and important information which can be updated obeying a constraint of the age of information (AoI) \cite{Sun2021}. With the centralized database from sensing layer and global networking information, the cybertwin can training communication and networking optimization models for services with on-demand access, where the artificial intelligence (AI) techniques can provide effective and flexible supports. Particularly, the information exchange between different DTs provides a global perspective for any local service, which enables improved resource sharing and utilization. By the cybertwin, we assume all the complex decisions made by DTs and the mobile agents and physical entity just execute its resource corresponding orchestration strategy and service function chain, thus the resource consumption of local mobile agents and access points can be significant reduced. 

\textbf{Resource layer:} Multi-source heterogeneous resources can be scheduled by the orchestration strategy from the cybertwin and we assume a reserved resource in the resource pool to ensure the system operation running and prepare for urgent services or response to abnormal faults. 

\textbf{Service layer:} We consider communication and computation services in the service layer.
Combining the sensed information and communication request, the communication system is designed with compatible transmission and access schemes which can support differentiated the communication requirement by flexibly scheduling communication resource. Particularly, due to random service requests and uncertain interference, discontinuous and available spectrum resources require a fusion scheme of broad and narrow band to improve the spectral efficiency, and an adaptive communication coverage is controlled by the cybertwin and executed by mobile RAN entities in SAGIN. For the computation services, the sensed information including the appropriate objects for task offloading and migration are reported to the cybertwin and a cooperative scheme can be realized. Besides, with analyzing the sensed information and current networking status, a computation fusion model can be directed to reduce the latency.

\section{Open Issues and Challenges}
\begin{center}
	\textbf{AI-based Networking Decision and Optimization in Cybertwin-enabled SAGIN }
\end{center}

In the cybertwin-enabled SAGIN, to intelligently and efficiently manage such multi-source heterogeneous network, the cybertwin uploads and updates various function models of decision and optimization which maps different services. Particularly, the DTs in SAGIN cybertwin hold many kinds of data, i.e., the environment information, the networking state information, and the service requested KPI, and it is expected immediately and accurately reflect the physical space of SAGIN. Considering the service diversity, the great variety and large volume of data, and the dynamic topology of physical SAGIN. it is challenging to effectively and efficiently manage the network. Whereas, to handle such challenges, AI-based approaches have the great potential in networking decision and optimization\cite{Jiang2020}.

\textit{Networking decision:} Based on the data warehouse created in the cybertwin, the function models of networking analysis, prediction, maintenance, etc. can be trained and updated by AI algorithms for a specific application, which achieves the model-decision making and decouples the control and service layers of physical SAGIN. Particularly, due to the multi-tier heterogeneous RANs in SAGIN, the access control decision is challenging to realize the global optimization by using traditional random or fixed access strategies. However, with the specific scenario information and the networking behavior learning, AI-based access control model can be realized in the cybertwin space and the model-decision can be delivered to the corresponding physical entity expediently.  

\textit{Networking optimization:} Although we have developed an integrated cloud-edge-end resource pool in SAGIN based on cybertwin, however, the resource from different physical network providers shows differentiated characteristics, which is leveraged to be constrained on specific indicators, i.e., delay, bandwidth, power, and computation, etc. Therefore, it is challenging to orchestrate and optimize the heterogeneous resources in SAGIN. To fulfill the on-demand resource allocation, the resource classification can be first conduct in the cybertwin according to the constrained KPIs. By information exchange among DTs, specific networking optimization models can be trained by AI-based approaches with using global resources. Particularly, these optimization model can be pre-generated through simulations with historical data in the cybertwin space, which is then mapped into the physical SAGIN entities to execute a networking optimization.     
  
\begin{center}
	\textbf{FL-based Cross-layer Privacy and Security in Cybertwin-enabled SAGIN}
\end{center}

We consider inter-DT communications among the DTs of SAGIN entity in the SAGIN cybertwin, and thus the cybertwin provides an information exchange platform through the meeting of DTs for users and entities in different physical networks. 
However, it is hard to establish an uniform confidential mechanism among the DTs to preserve the private data of host from being exposed to adversaries, since the DT is possible to be falsely impersonated. Thus, the physical SAGIN network is vulnerable to be attacked if the DT security could not be ensured.  
Moreover, since the data hold in a DT is the digital asset which has significant research values and economic benefits, it can be traded in the digital market. The host of DT will experience the economic loss and the privacy leak when its DT is confused.

Particularly, as a distributed ML approach, FL has been introduced to address the issue of privacy protection, where the ML model can be trained at local end and which is then sent 
to the edge/cloud server\cite{Wei2020}. It sends the model updates rather than raw data to the server for aggregation. However, for the model transmissions via wireless communications, it is vulnerable to be eavesdropped. 

To address the challenges of privacy and security in cybertwin-enabled SAGIN, we direct a potential research issue of FL-based cross-layer privacy and security in cybertwin-enabled SAGIN. Particularly, the local private data can be trained at the DT of this end and an secure transmission should be ensured in physical layer when the local model is uploading and the aggregated model is downloading.

\section{A Case Study}
In this section, we conduct a case study of cybertwin-enabled deep integration in SAGIN.
Particularly, considering the heterogeneity of SAGIN, heterogeneous mobile agents served by different RANs of SAGIN, i.e., satellite network, aerial network, and terrestrial network, only focus on its local resource and it is difficult to join in a comprehensive training crossing different tiers. To investigate the cybertwin-enabled comprehensive training in SAGIN, we train a classification model based on the MNIST data set in the cybertwin space of SAGIN. Specifically, we consider a simple system model of cybertwin-enabled SAGIN where 
the physical RANs consisting of a LEO satellite, an UAV, and a terrestrial BS. There are multiple user agents connecting to satellite, UAV, and BS, respectively. 
Particularly, we conduct a simulation to evaluate the model training with FL in cybertwin-enabled SAGIN and the system parameters are specifically set as follows.

In satellite network, the height of satellite orbit is 600 Km, the maximum beam gain is 46 dB, and the 3-dB angle of satellite beam is set to 0.4$^\circ $. The rain attenuation parameters of satellite-terrestrial channel are set to -3.152 dB and 1.6 \cite{Yin2021}. 
In the UAV-based aerial network, the horizontal distance from UAV to the boundary of coverage is 100 m, the UAV altitude is 100 m, and Rician fading model is adopted as the small-scale fading. The UAV transmit power is set to 15 dB. The number of UAV downlink transmit antennas is four. For the terrestrial network, the BS is equipped with four transmit antennas and the transmission power is set to 40 dB. Besides, there are five users connected to satellite, UAV, and terrestrial BS, respectively. 
In addition, the learning model contains a flatten layer, a dense layer with 128 units, a dropout layer with a scale of 0.2, and an output layer with 10 units. 
The data set held by each user contains 200 samples.
For the comparison, we also evaluate the performance of independently performing FL in satellite network, UAV-based network, and terrestrial network, respectively.

\begin{figure}[!h]
	\centering
	\includegraphics[width=0.5\textwidth]{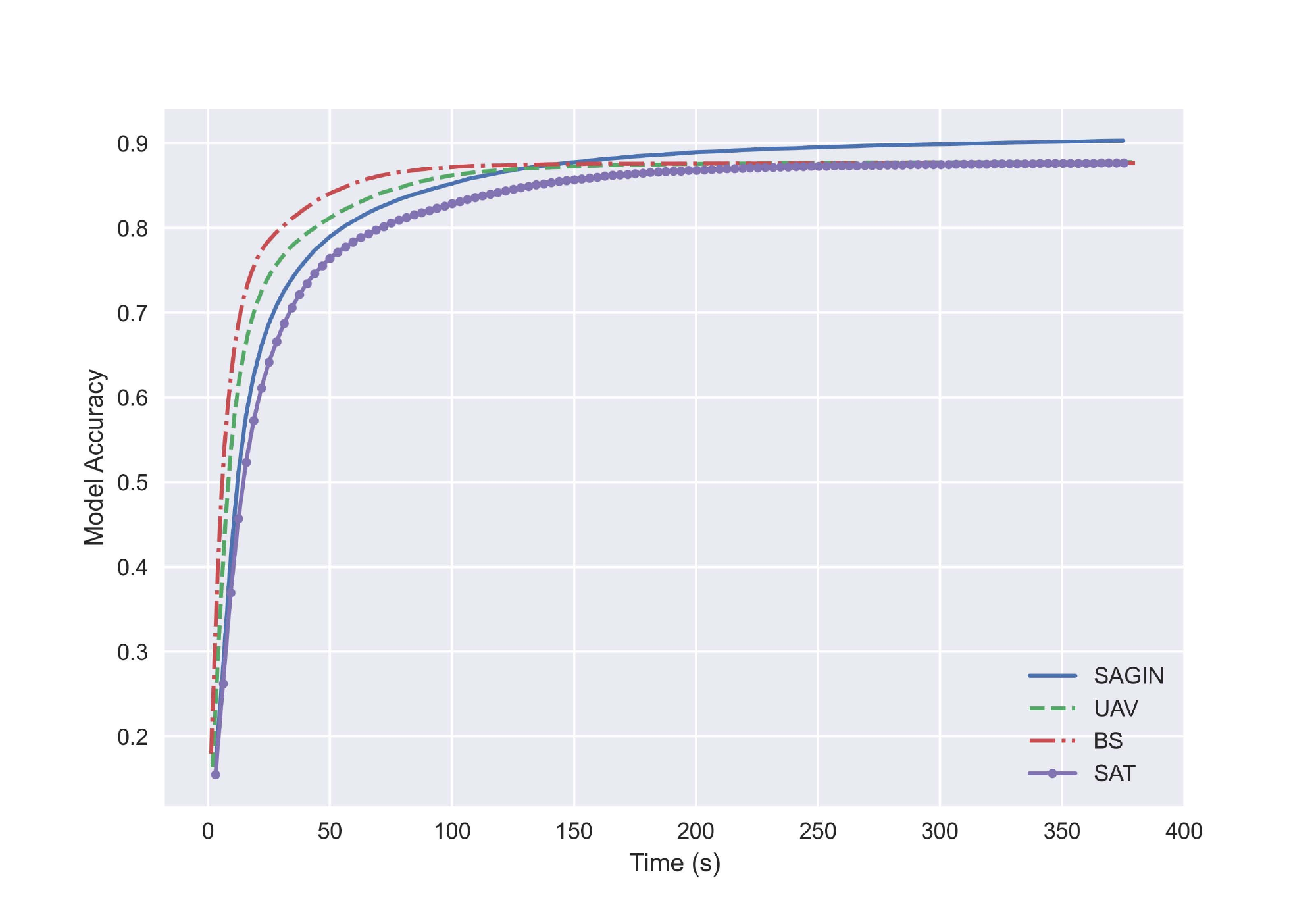}
	\caption{The model accuracy of FL in cybertwin-enabled SAGIN}
	\label{Simu}
\end{figure}

Fig. \ref{Simu} shows the model accuracy of FL in cybertwin-enabled SAGIN.
From Fig. \ref{Simu}, it can be seen that 
the FL in terrestrial network shows the fastest convergence speed (convergence at about 100 s), followed by the UAV-based network (convergence at about 150 s), and that in satellite network and SAGIN are the slowest (convergence at about 300 s).
 This is because the delay of communication link in satellite network is larger due to its long-distance transmission link, followed by the UAV-based network, then is that in the terrestrial network. Thus the date updates slower in Sat-DT than UAV-DT and BS-DT,
 which leads to a longer time for training the model with Sat-DT. 
Moreover, due to the straggler effect, the convergence speed is determined by the segment with slowest convergence speed, therefore the model training of FL in SAGIN converges slowly. In addition, it is observed that the higher model accuracy can be achieved when performing FL in the cybertwin of SAGIN. This is because all the heterogeneous mobile agents can participate in training in the cybertwin of SAGIN, which converts the local training to a global training, thereby the model accuracy increases.
\section{Conclusion}
In this paper, to support the on-demand service and improve the networking efficiency, a cybertwin-enabled SAGIN architecture has been presented to reach the deep integration of multi-tier heterogeneous networks and envision the ubiquitous intelligence towards 6G.  
Particularly, with the cybertwin, we have presented three categories of critical technologies in SAGIN, i.e., cybertwin-based multi-source heterogeneous network integration, cybertwin-based integrated cloud-edge-end, and cybertwin-based integrated sensing-communication-computing. They can support the implementation of cybertwin-enabled  SAGIN with improved networking efficiency and service experience. 
In addition, we have presented and discussed two open research issues, and the potential solutions have also been directed accordingly by the AI-based approach and FL respectively. Finally, a case study of model training in cybertwin-enabled SAGIN has been conducted through FL and the result shows an increased model accuracy by the cybertwin-enabled networking integration in SAGIN.    


%



%
%

\ifCLASSOPTIONcaptionsoff
  \newpage
\fi



\bibliographystyle{IEEEtran}
\bibliography{IEEEabrv,cybertwin}
%

%








\end{document}